\documentclass{elsart}

\usepackage{epsfig}

\usepackage{amssymb}

\newcommand{\fig}[1]{Fig.~\ref{#1}}

\begin{document}

\begin{frontmatter}



\title{Heavy quark jet quenching with collisional plus radiative energy loss and path length fluctuations}


\author[label1]{Simon Wicks}
\ead{simonw@phys.columbia.edu}
\author[label1]{William Horowitz}
\ead{horowitz@phys.columbia.edu}
\author[label2]{Magdalena Djordjevic}
\ead{magda@mps.ohio-state.edu}
\author[label1]{Miklos Gyulassy}
\ead{gyulassy@phys.columbia.edu}

\address[label1]{Columbia University, Department of Physics, 538 West 120th Street, New York, NY 10027}
\address[label2]{Department of Physics, The Ohio State University, 191 West Woodruff Avenue, Columbus, OH 43210}

\begin{abstract}
With the QGP opacity computed perturbatively and with the global entropy constraints imposed by the observed $dN_{ch}/dy\approx1000$, radiative energy loss alone cannot account for the observed suppression of single non-photonic electrons. We show that collisional energy loss, which previously has been neglected, is comparable to radiative loss for both light and heavy jets and may in fact be the dominant mechanism for bottom quarks. Predictions taking into account both radiative and collisional losses significantly reduce the discrepancy with data. In addition to elastic energy loss, it is critical to include jet path length fluctuations to account for the observed pion suppression.
\end{abstract}

\end{frontmatter}
\section{Introduction}
Light quark and gluon jet quenching~\cite{phenix_pi0} observed via $\pi,\eta$
suppression in Cu+Cu and Au+Au collisions at
$\sqrt{s} = 62-200$~AGeV at the Relativistic Heavy Ion Collider (RHIC)
has been remarkably consistent thus far with
predictions. However,
recent non-photonic single electron data~\cite{elec:PHENIX,elec:STAR}, which
present an indirect probe of heavy quark energy loss, have
significantly challenged the underlying assumptions of the jet
tomography theory.  A much larger
suppression of electrons than predicted was observed in the
$p_T\sim 4-8 $ GeV region
(see \fig{fig:eandpiRAA}). These data falsify the assumption that heavy quark
quenching is dominated by radiative energy loss when the bulk QCD
matter parton density is constrained by the observed $dN_{ch}/dy\approx
1000$ of produced hadrons.

\begin{figure}[!th] 
\centering
\epsfig{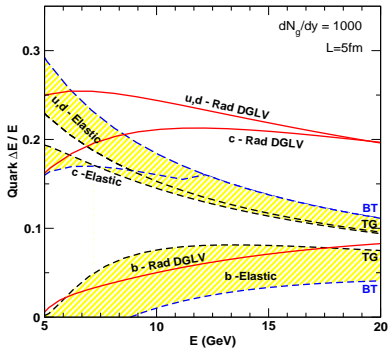} 
\caption{ 
\label{fig:deltaE} 
Average $\Delta E/E$ for $u,c,b$ quarks as a function of $E$ in a Bjorken
expanding QGP. 
Radiative DGLV first order energy loss is compared to elastic parton
energy loss in TG or BT approximations.
}
\end{figure}

This discrepancy between radiative energy loss predictions and current data and recent papers motivated us to
revisit the assumption that pQCD elastic energy
loss is negligible compared to radiative energy loss. In
some earlier studies, the elastic energy
loss was found to be $dE^{el}/dx\sim 0.3-0.5 $ GeV/fm, which was
erroneously considered to be small compared to the several GeV/fm expected
from radiative energy loss. In \fig{fig:deltaE} we see that above $E>10$ GeV the light and charm quarks have 
elastic energy losses smaller but of the same order of magnitude as the inelastic losses. Due to the 
large mass effect, both radiative and elastic energy losses remain significantly smaller for
bottom quarks than for light quark and charm jets, but the elastic loss can now be greater than inelastic up to $\sim15$GeV. 
The uncertainties from the Coulomb log, as illustrated by the difference between the TG and BT 
lines~\cite{Thoma:1990fm,Braaten:1991jj}, are largest for the heaviest b quark:  as they are not ultrarelativistic, 
the leading log approximation breaks down in the jet energy range accessible at RHIC.

\section{RHIC predictions and uncertainties}
We present a calculation of jet suppression using the model explained in \cite{Wicks:2005gt}. We assume initial $dN_g/dy=1000$ and a fixed
coupling, $\alpha_s=0.3$. The main difference from the previous 
calculation~\cite{Djordjevic:2005db} 
is the inclusion of two new physics components in the  
energy loss probability $P(E_i \rightarrow E_f )$.
First, $P(E_i \rightarrow E_f)$ is generalized to include  
both elastic and inelastic energy loss and their fluctuations. The second major
change is that we now take into account geometric path length fluctuations. The geometric path 
averaging used here is similar to that used elsewhere, but 
the inclusion of elastic energy loss together with path fluctuations
in more realistic geometries was not considered.

\begin{figure}[!ht]
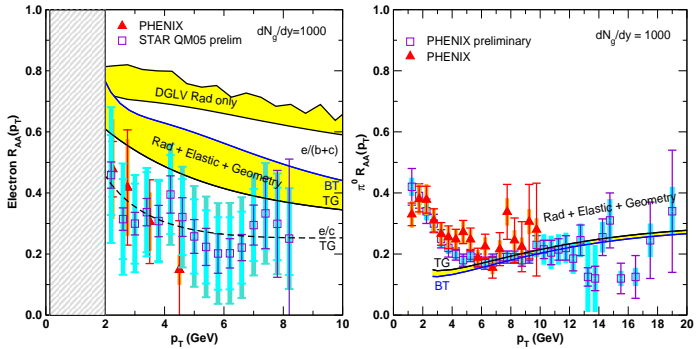
 
\centering
$\begin{array}{cc}
\epsfig{file=plot_rhic_raa_el_l45_l55.eps,height=1.8in,width=1.8in,clip=,angle=0} 
\epsfig{file=plot_rhic_raa_pi_l35_l5.eps,height=1.8in,width=1.8in,clip=,angle=0} 
\end{array}$
\caption{\label{fig:eandpiRAA} 
The suppression factor, $R_{AA}(p_T)$, of non-photonic electrons (left) and pions (right) in central Au+Au reactions at
200~AGeV are compared to data. For electrons, the upper yellow band~\cite{Djordjevic:2005db}
takes into account radiative energy loss only, using a fixed $L=6$ fm; the lower yellow band is the new prediction.
The dashed curves illustrate the lower extreme
of the uncertainty from production, by showing the radiative plus elastic prediction with bottom quark jets neglected.
}
\end{figure}

The results for the suppression of non-photonic single electrons are shown in the upper plot in \fig{fig:eandpiRAA}. 
As emphasized in~\cite{Djordjevic:2005db}, any proposed energy loss mechanisms must also be checked for consistency with the extensive pion
quenching data~\cite{phenix_pi0}, for which preliminary data
now extend out to $p_T\sim 20$ GeV. This is also shown in \fig{fig:eandpiRAA}.

\begin{figure}[th]
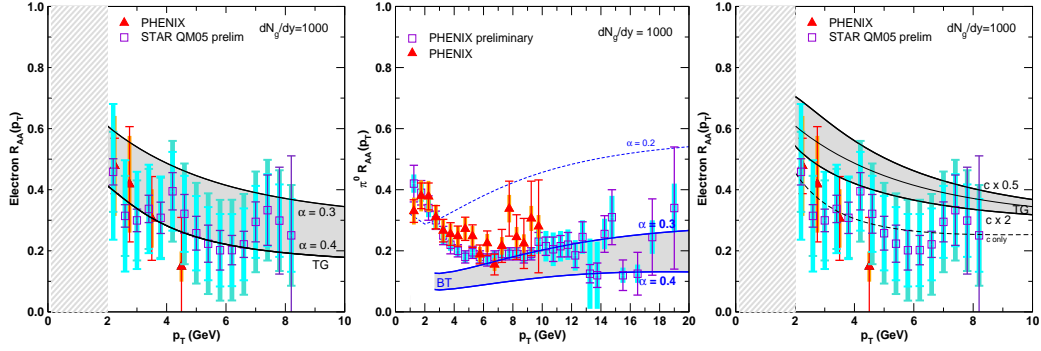

\centering
$\begin{array}{ccc}
\epsfig{file=Varyingalpha.eps,height=1.8in,width=1.8in,clip=,angle=0} 
\epsfig{file=Varyingalpha_pi.eps,height=1.8in,width=1.8in,clip=,angle=0} 
\epsfig{file=Varyingctob.eps,height=1.8in,width=1.8in,clip=,angle=0}
\end{array}$
\caption{ 
\label{fig:alpha} 
The variation in $R_{AA}(p_T)$ predictions is shown for change in the fixed coupling $\alpha_s$ and for variation in the charm to bottom ratio
in the production spectra.
}
\end{figure}

It is important to examine the theoretical uncertainties involved in these predictions. Uncertainty in the leading log approximation has already been shown and two other sources are illustrated in \fig{fig:alpha}. The radiative and elastic energy losses are strongly
dependent on the coupling. To estimate the uncertainty involved from this approximation, the results of varying 
$\alpha_s$ are shown. While increasing fixed $\alpha_s$ to 0.4 improves the fit to the electron data, this then overpredicts the pion
quenching.

The ratio $R_{AA}$ is not sensitive to the scaling of all cross-sections by a constant. However, the electron $R_{AA}$ is sensitive to any uncertainty in the relative contribution of charm and bottom jets~\cite{Armesto:2005mz}. The result of changing the charm to bottom production ratio by a constant is shown, as well as the lower bound extreme of electrons from charm jet only.

\section{Conclusion}
The elastic component of the energy loss cannot be neglected when considering pQCD jet quenching.
While the results are encouraging, further improvements will be required before stronger conclusions can be drawn.
It will be important to deconvolute the charm and bottom contribution to the electrons, so direct measurement of $D$ spectra
will be essential. On the theoretical side, further work on the deconvolution of coherence and finite time effects as well
as implementing a running coupling will significantly reduce the theoretical uncertainties in the predictions.



\end{document}